\documentclass{PoS}

\title{Low Luminosity Radio Loud Active Galactic Nuclei}

\ShortTitle{Low Luminosity AGNs}

\author{\speaker{Gabriele Giovannini}$^{1,2}$ and {Marcello Giroletti}$^{2}$\\
\llap{$^1$}Astronomy Department - Bologna University\\
     via Ranzani 1, 40127 Bologna, Italy\\
\llap{$^2$}INAF - Istituto di Radioastronomia\\
     via Gobetti 101, 40129 Bologna, Italy\\
  E-mail: \email{ggiovann@ira.inaf.it}, \email{giroletti@ira.inaf.it}}

\abstract{ I review observational properties of low power radio loud AGN. High
resolution VLBI observations allow the estimate of the jet velocity and
orientation with respect to the line of sight and the determination of the
Doppler factor.  These data reveal rich structures, including two-sided jets
and secondary components.
New results on 1144+35, a giant radio source with superluminal motion are
shown in detail.
}

\FullConference{The 8th European VLBI Network Symposium on New Developments in VLBI Science and Technology
                and EVN Users Meeting  \\
		September 26-29 2006\\
		Torun, Poland}

\begin{document}

\section{Introduction}

The study of parsec scale properties of radio galaxies is crucial to derive
physical properties of the central regions of the Active Galactic Nuclei (AGN)
and provides the observational basis of Unified Models. A comparison
of parsec (pc) and kiloparsec (kpc) scale morphology and properties
is the key to understand the origin and evolution of extended radio sources.

To investigate the properties of radio galaxies at pc resolution it is 
important to 
select sources from low frequency radio catalogues 
where the source
properties are dominated by the unbeamed extended emission
and are not affected by observational biases related to orientation effects.
To this aim,
we undertook a project of observations of a complete sample of radio
galaxies selected from the B2 and 3CR catalogs
with z $<$ 0.1 (i.e. no constrain
on the core flux density): the {\it Bologna Complete Sample} (BCS~~ 
\cite{gio05}).
This sample consists of 95 sources. At present 60 on 95 sources have
been studied with VLBI observations. We observed as a first step sources
with an arcsecond core flux density $\sim$ 10 mJy or more at 5 GHz and are
observing now with phase reference observations sources with a core flux 
density in the range 1 to 10 mJy (see e.g. Fig. 1).

\section{Sources Morphology}

Parsec scale structures are mostly one-sided because of relativistic Doppler
boosting effects, however a large number of sources with two-sided jets
have been found.

\begin{figure}[H]
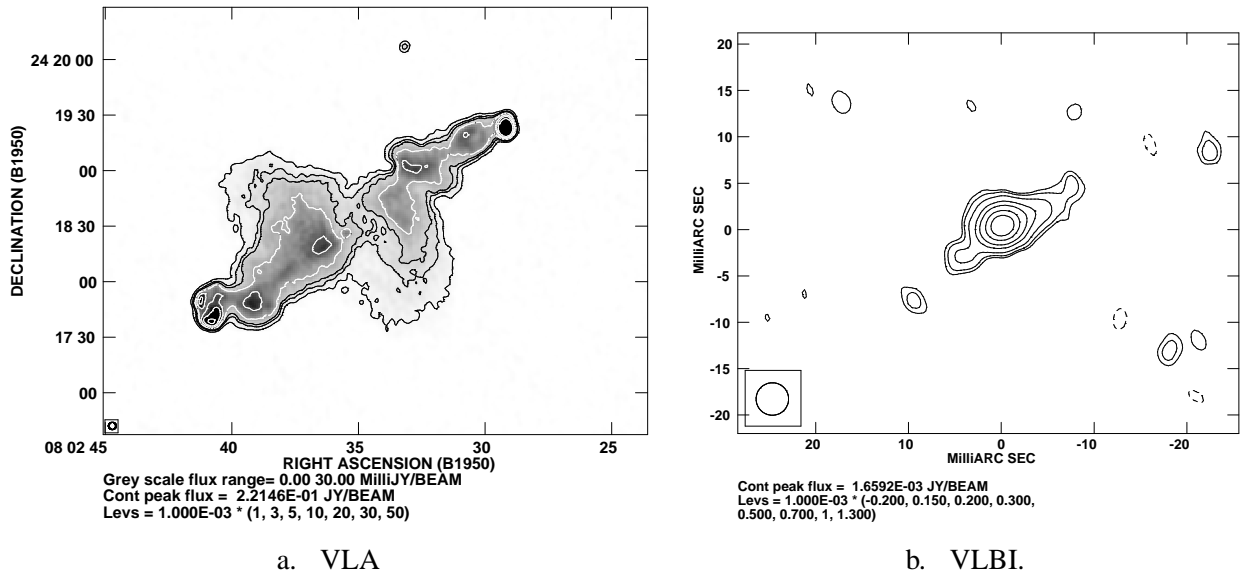

\centerline{%
\begin{tabular}{c@{\hspace{0.5pc}}c}
\includegraphics[width=3.5in]{3C192.ps} &
\includegraphics[width=3.0in]{3c192.vlbi.ps} \\
a.~~ VLA & b.~~  VLBI.
\end{tabular}}
\caption{VLA image of the kpc scale structure of 3C192 from
http://www.jb.man.ac.uk/atlas (a).
Phase reference VLBA image of the core of 3C192 (b).}
\label{fig5}
\end{figure}

Among observed sources 18 show a two-sided pc scale structure corresponding
to $\sim$ 30\%. We note that in previous surveys of large samples selected
at high frequency as the Caltech survey the percentage of two-sided structures
is only of 4.6\%.
The difference between the percentage of symmetric sources in the
present sample and in previous samples is naturally explained in the
framework of unified scheme models by the fact that present
sources have been selected at low frequency, show relatively faint cores, 
and are therefore less affected by
orientation bias.  

In most sources we find a good agreement between the pc and kpc-scale
structures. This result 
supports the suggestion that the large distortions detected in BL-Lac
sources are due to small intrinsic bends amplified by the small angle
of the BL-Lac jets with respect to the line-of-sight.

\section{Jet Kinematics}

\subsection{Proper Motion}

Many AGNs contain compact radio sources with different components which
appear to move apart. Multi epoch studies of these sources allow a direct
measure of the apparent jet pattern velocity ($\beta_a$c). 
From the measure of $\beta_a$ we can
derive constraints on $\beta_p$ and $\theta$ where $\beta_p$c is the intrinsic
velocity of the pattern flow and $\theta$ is the jet orientation with respect
to the line of sight:

\begin{equation}
\beta_p = \beta_a/(\beta_a cos\theta + sin\theta)
\end{equation}

A main problem is to understand the difference between the bulk and pattern
velocity. In few cases where proper motion is well defined and the bulk
velocity is strongly constrained, there is a general agreement between the
pattern velocity and the bulk velocity (see e.g. NGC 315 \cite{cot99},
and 1144+35, here). However, in the same source we can have different
pattern velocities as well as stationary and high velocity moving structures.
Moreover, we note that in many well
studied sources the jet shows a smooth and uniform surface brightness and
no (or very small) proper motion (as in the case of Mrk 501,
\cite{gir04}).

\subsection {Bulk Velocity}

Assuming that the jets are intrinsically symmetric we can use relativistic
effects to constrain the jet bulk velocity $\beta$c and orientation with
respect to the line of sight ($\theta$), as discussed e.g. in \cite{gio04}.

I will discuss here only the jet -- counter jet (cj) brightness ratio
and the core dominance since they are the mostly used methods.

\begin{itemize}

\item
Assuming that the jets are intrinsically symmetric we can use the observed
jet to cj brightness ratio R to constrain the jet
bulk velocity $\beta$c
and its orientation with respect to the line of sight:

\begin{equation}
R = (1+\beta cos\theta)^{2+\alpha} (1-\beta cos\theta)^{-(2+\alpha)}
\end{equation}

where $\alpha$ is the jet spectral index (S($\nu$) $\propto$ $\nu^{-\alpha}$).

\item
The core radio emission measured at 5 GHz, at arcsecond resolution is dominated
by the Doppler-boosted pc-scale relativistic jet.
The source radio power measured at low frequency (e.g. 408 MHz), instead,
is due
to the extended emission, which is not affected by Doppler
boosting. At low frequency the observed core radio emission is
not relevant since it is mostly self-absorbed.
Given the existence of a general correlation between the core and total
radio power discussed in  \cite{gio01},
we can
derive the expected intrinsic core radio power from the {\it unboosted}
total radio power using the estimated best fit correlation (continuum line in
Fig. 2):

\begin{equation}
log P_c = (0.62\pm0.04) log P_t + (7.6 \pm 1.1)
\end{equation}

The comparison between the expected intrinsic core radio
power and the observed core radio power will give constraints on the jet
velocity and orientation (\cite{gio01}).
\end{itemize}

The large dispersion in the core
radio power visible in Fig. 2 is expected because
of the strong dependance of the observed core radio power on $\theta$ and
$\beta$.

\begin{figure}
\centerline{
\includegraphics[width=24pc]{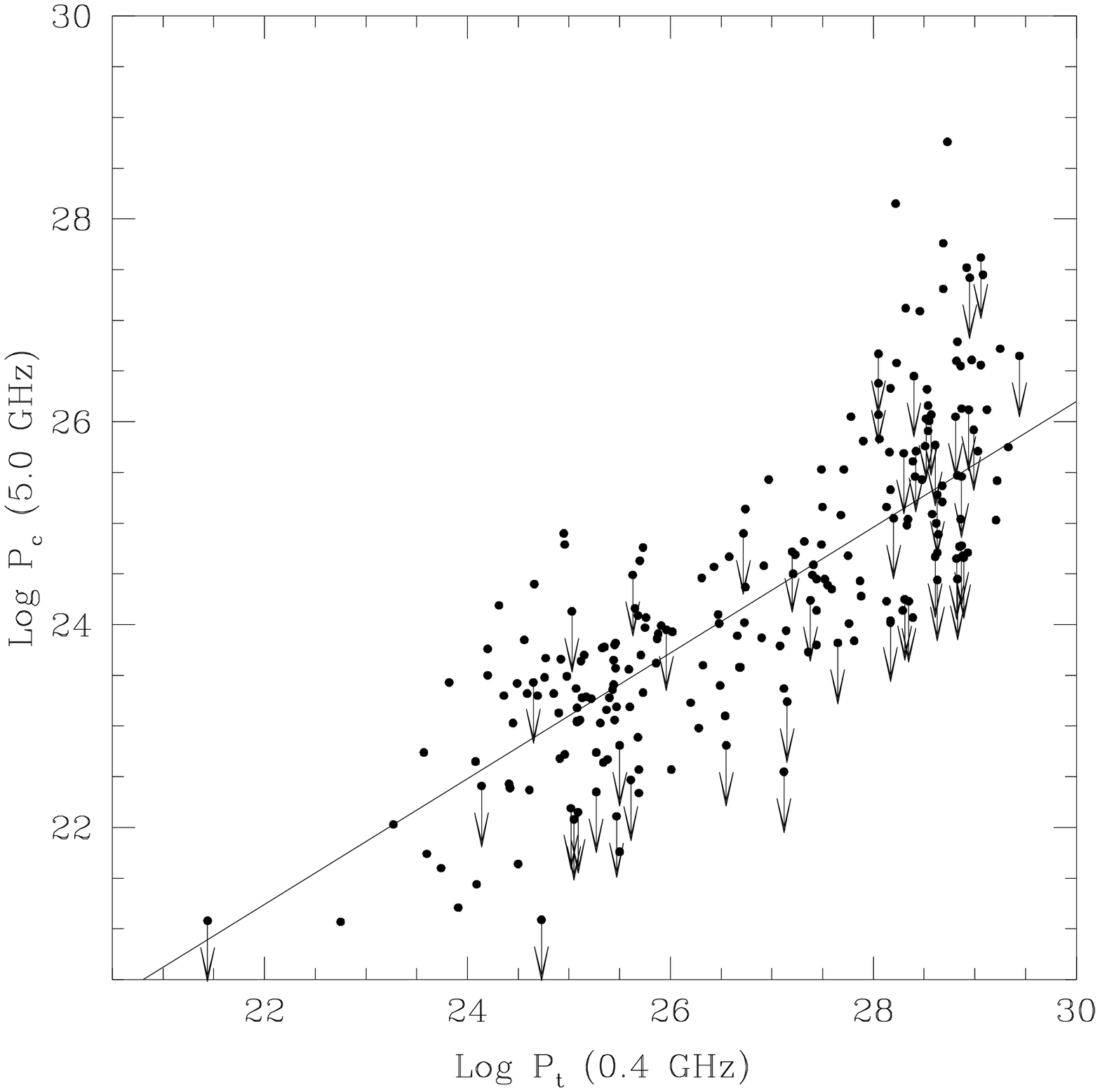}}
\caption{Arcsecond core radio power at 5 GHz (P$_c$) versus total radio power
at 0.4 GHz (P$_t$) for B2 and 3CR radio sources (see \cite{gio88}). 
Arrows are upper limits when a nuclear emission was
not detected. The line is the best fit of the data corresponding to
a soource orientation with respect to the line of sight of 60$^\circ$.}
\label{fig2}
\end{figure}

From the data dispersion, assuming that no selection effect is present in
the source orientation ($\theta$ = 0$^\circ$ to 90$^\circ$),
we can derive that the
jet Lorentz factor $\Gamma$
has to be in the range 3 to 10 otherwise we should observe a smaller or
larger core radio power
dispersion.

\section{Results}

To derive statistical properties of radio jets on the pc scale, we
used all observational data for the 60 sources in our sample with VLBI data.
We found that in all sources pc scale jets move at high velocity. No
correlation has been found between the jet velocity and  the core or total
radio power.
Highly relativistic parsec scale jets
are present regardless of the radio source power. Sources with a different kpc
scale morphology, and
total radio power have pc scale jets moving at similar velocities.

We used the estimated jet $\beta$ and $\theta$ to derive
the Doppler factor $\delta$ for each source,
and the corresponding intrinsic core radio power (assuming $\alpha$ = 0):

\begin{equation}
P_{c-observed} = P_{c-intrinsic} \times \delta^2
\end{equation}

We found a good correlation between  P$_{c-intrinsic}$ and P$_{t}$ with a
small dispersion
since plotting P$_{c-intrinsic}$, we removed the spread due to
the different orientation angles (Fig. 3).
\begin{figure}[H]
\centerline{%
\begin{tabular}{c@{\hspace{0.5pc}}c}
\includegraphics[width=3.0in]{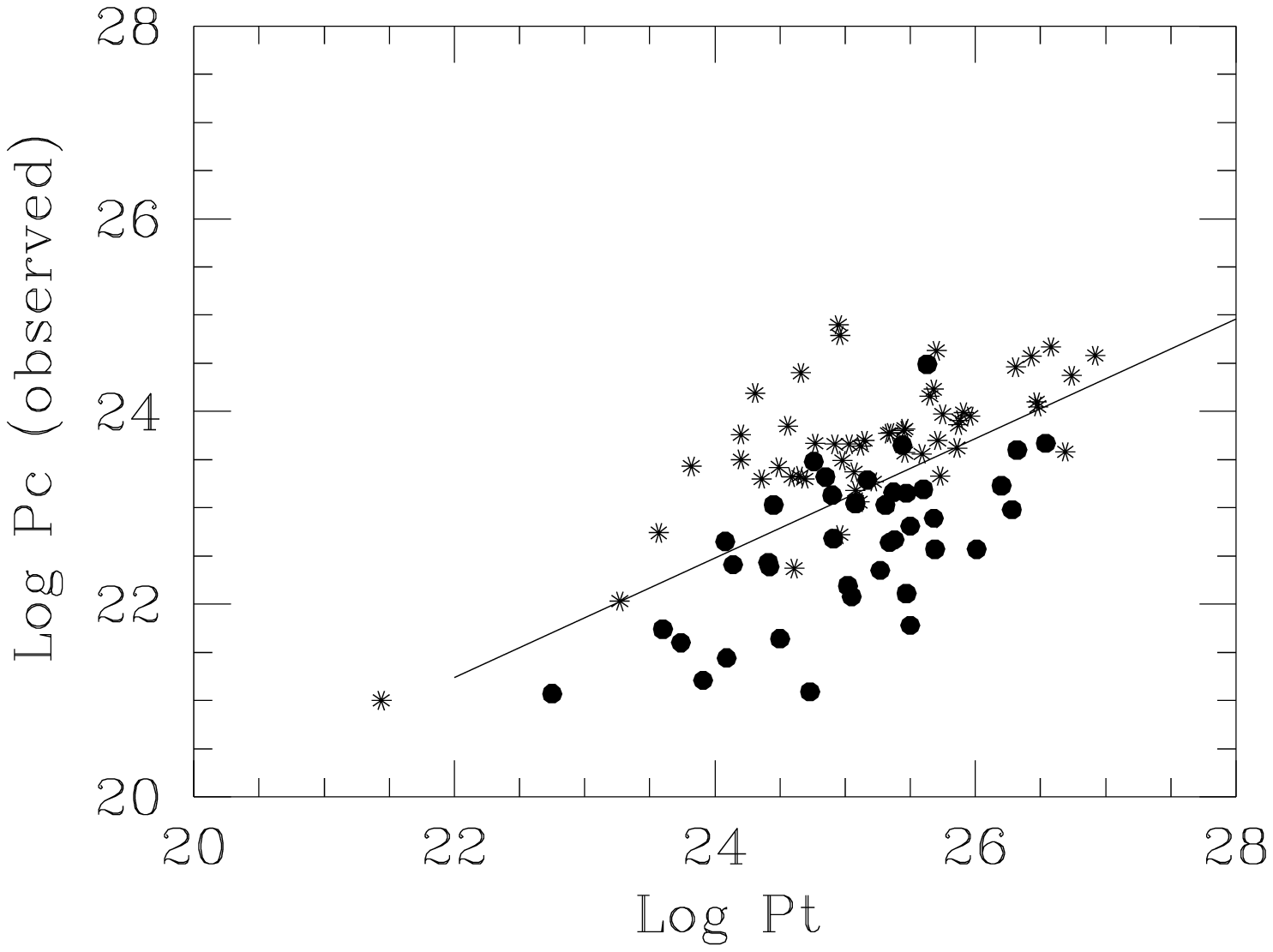} &
\includegraphics[width=3.0in]{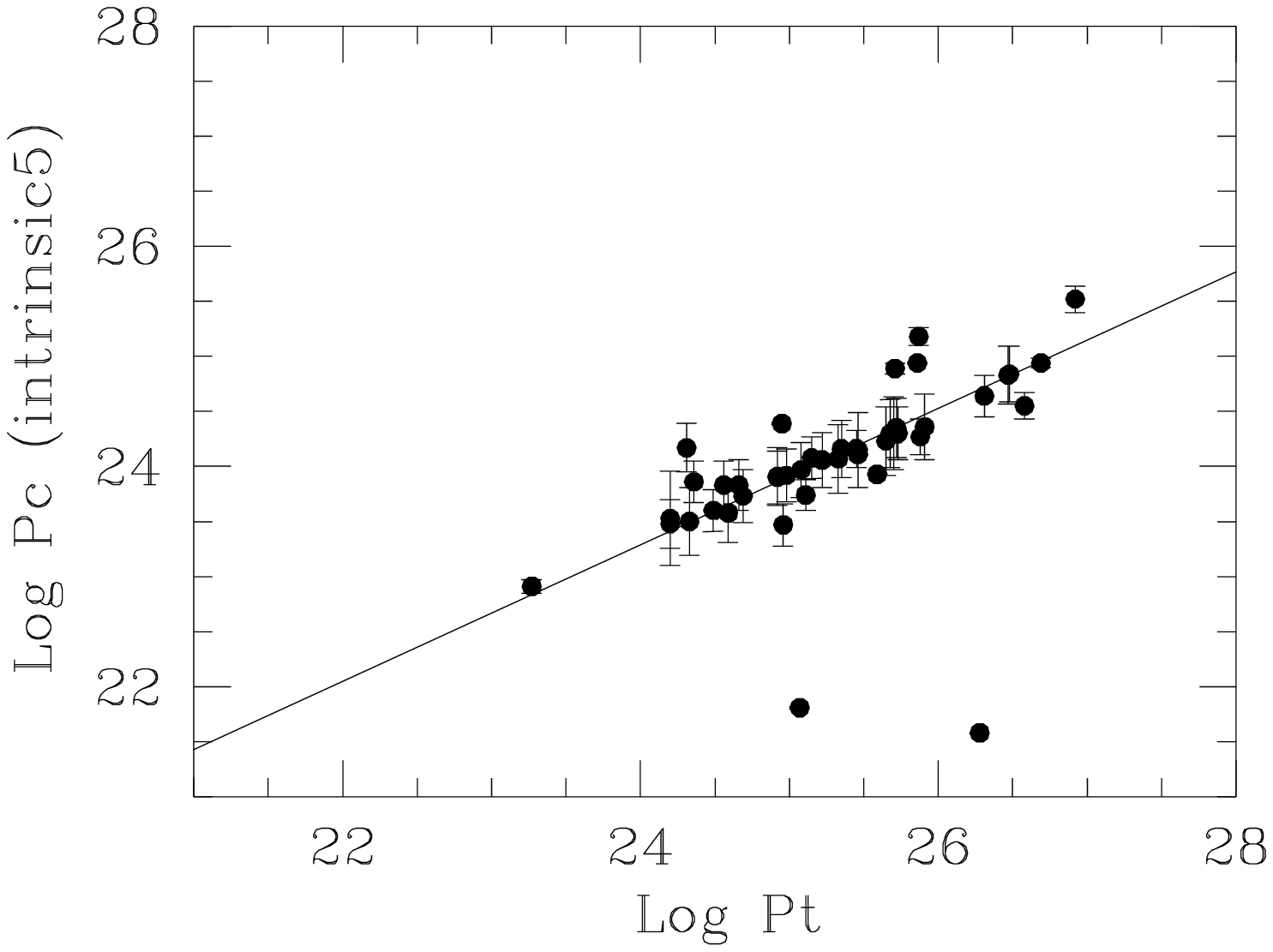} \\
a.~~ observed data & b.~~  intrinsic data.
\end{tabular}}
\caption{Observed core radio power versus total radio power for sources 
from the BCS with VLBI data (a).
 As (a) but with intrinsic core radio power versus total radio power (b).}
\label{fig3}
\end{figure}

These results are in agreement with the expectations from unified models.

\section{Low power radio cores}

The previous results are supported by the studies of many 
sources. However 
studying sources with a faint arcsecond scale radio core never studied
before in a large sample, we are having a relatively large number of peculiar
pc scale structures which interpretation and connection with the lage scale
structure it is not clear. 
The present low number of studied sources does not allow 
to distinguish between 
a general
property of sources with a mJy (or lower) radio core and the presence
of a few peculiar sources, possibly with evidence of restarted activity.

Here we present the radio image at kpc and pc resolution of 3C310.
A more detailed discussion will be presented and discussed in Giovannini et 
al. (in preparation) together with other peculiar structures in our sample 
(e.g. 1346+26 and 0836+29B).

At kpc scale 3C310 is an extended relaxed double source with a bright core 
and a
short jet in N direction (see Fig. 4). The lobes show a filamentary 
structure, moreover 
total flux density measures from the NED database show that 3C310 has a 
steep radio spectrum ($\alpha$ $>$ 1). These two evidences suggest that the
radio lobes are old, but a detailed study is not available in literature.

\begin{figure}
\centerline{
\includegraphics[width=20pc]{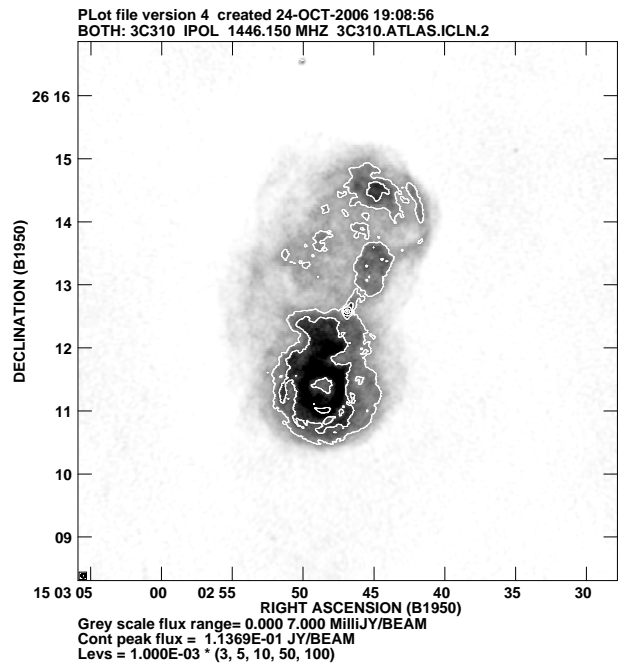}}
\caption{VLA Arcsecond radio image of 3C310 from 
http://www.jb.man.ac.uk/atlas.}
\label{fig2}
\end{figure}

As noted by \cite{mar99} for the optical image, this galaxy is 
flattened east-west, on an almost perpendicular direction to the kpc radio jet 
axis.

VLBI images (see also \cite{giz02}) show an extended structure 
elongated perpendicularly 
to the kpc scale radio structure and to the short jet visible in the VLA
image (see Fig. 5). The pc scale structure is very peculiar
since a dominant core structure is not visible even at 5 GHz.
With present data it is not possible to identify a core emission moreover
the extended emission is resolved and looks quite different from 
most pc scale jets detected in radio galaxies.
The low brightness and the absence of evident boosting effects suggest an 
orientation
near to the plane of the sky, but in this case the large difference in the 
position angle  
of the pc and kpc scale structure has to be real and not amplified by 
projection effects. More data at different resolution are necessary to 
understand the pc scale structure and its connection with the kpc scale radio
lobes.

\begin{figure}[H]
\centerline{%
\begin{tabular}{c@{\hspace{0.5pc}}c}
\includegraphics[width=2.0in]{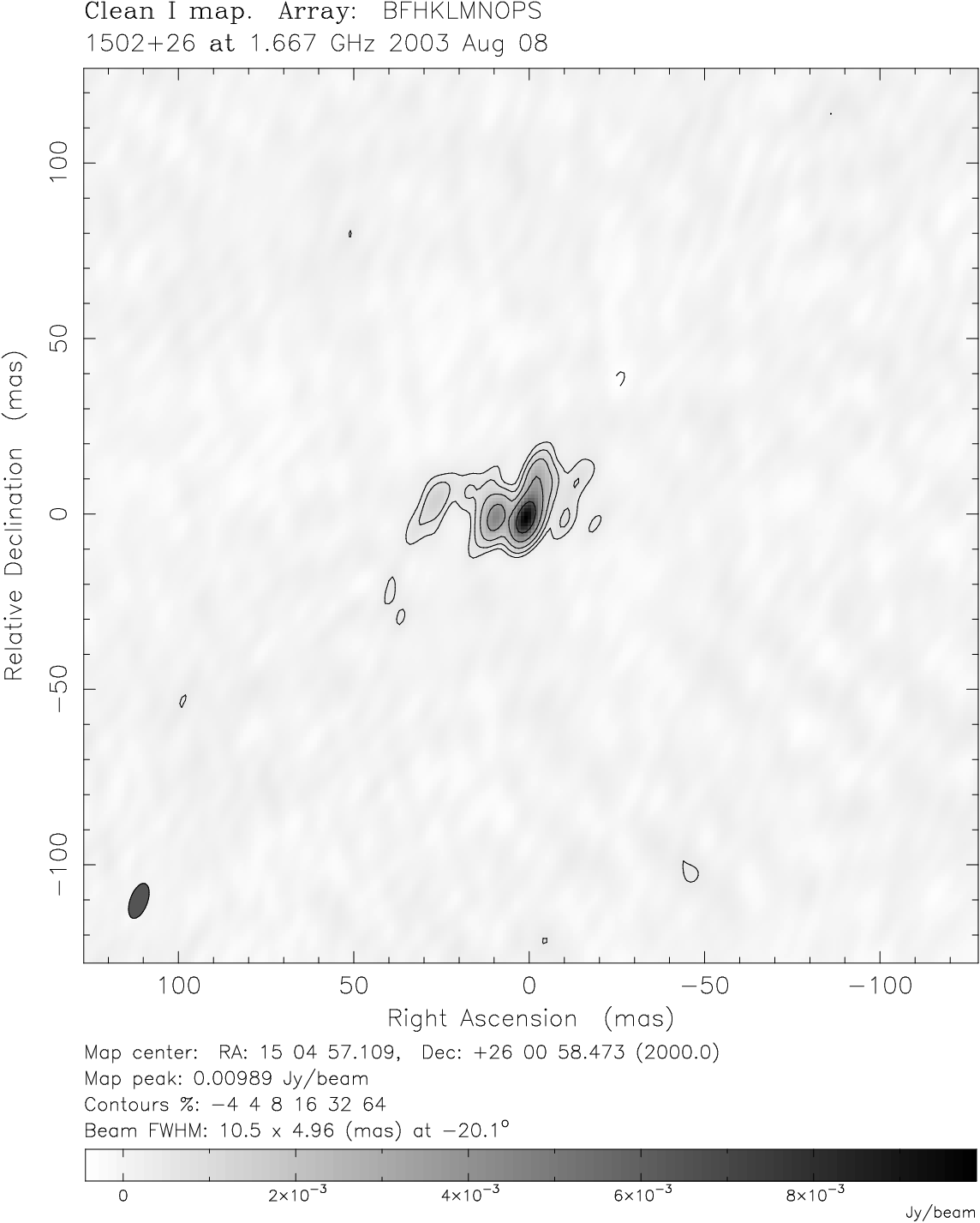} &
\includegraphics[width=3.0in]{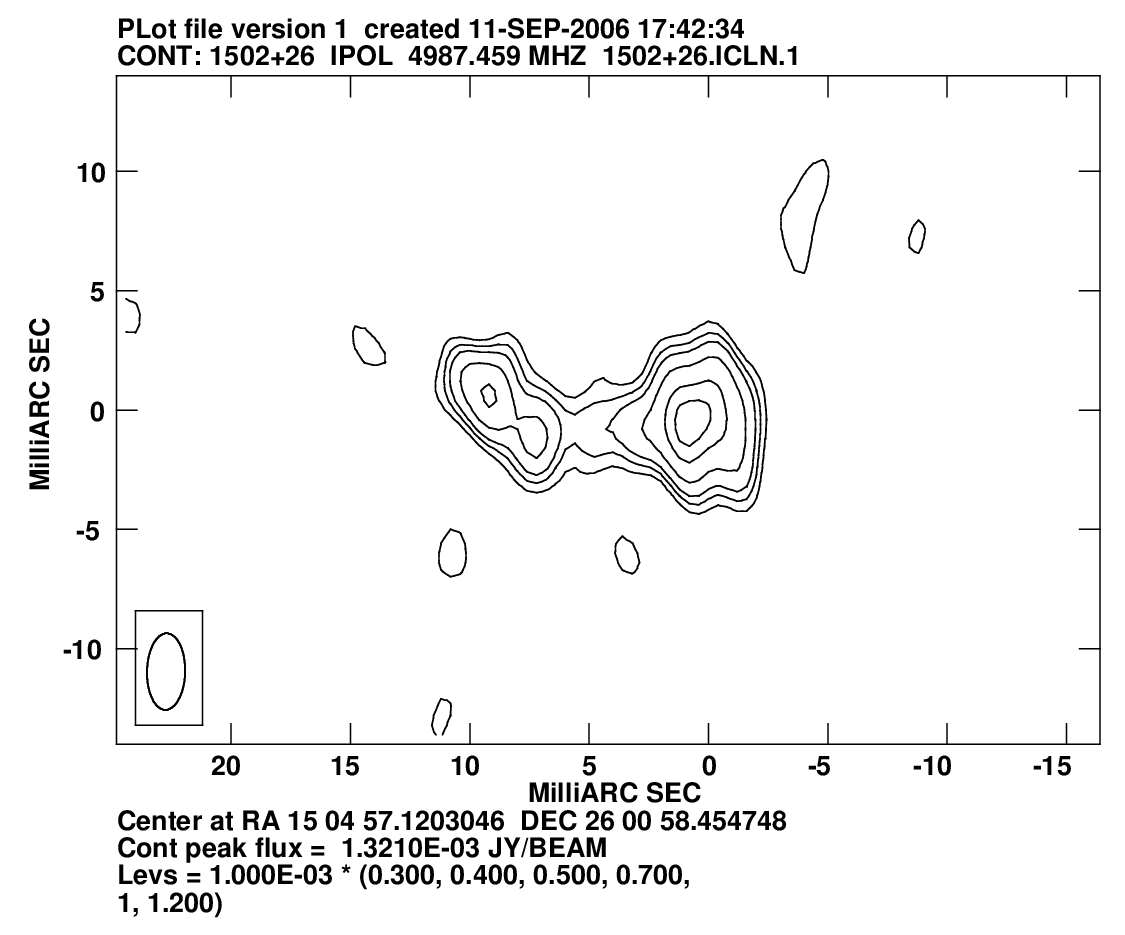} \\
a.~~  VLBI image at 1.4 GHz& c.~~ VLBI image at 5 GHz.
\end{tabular}}
\caption{
Phase reference VLBI images at 1.4 GHz(a) and at 5 GHz (b) of 3C310.}
\label{fig3}
\end{figure}

\section {1144+35}

The low power radio source B2 1144+35B was identified with
a faint (m$_{pg}$ = 15.7)
Zwicky galaxy (ZW186.48) in a medium-compact galaxy cluster at a redshift of
0.0630.

From the radio point of view, 1144+35 has a peculiar structure as discussed
in detail by \cite{gio99}.
New observations confirm the general structure discussed in \cite{gio99}.
The parsec scale structure (Fig. 6a) is resolved in
a nuclear source (C) with a short jet (D) and
counter-jet structure (E). At about 25 mas from the core
we have an extended jet like structure (A and B components) which is the
dominant structure in the VLBI images.
This jet structure is extended, clearly limb-brightened and connected to the
core by a low brightness emission.

\begin{figure}[H]
\centerline{%
\begin{tabular}{c@{\hspace{0.5pc}}c}
\includegraphics[width=3.0in]{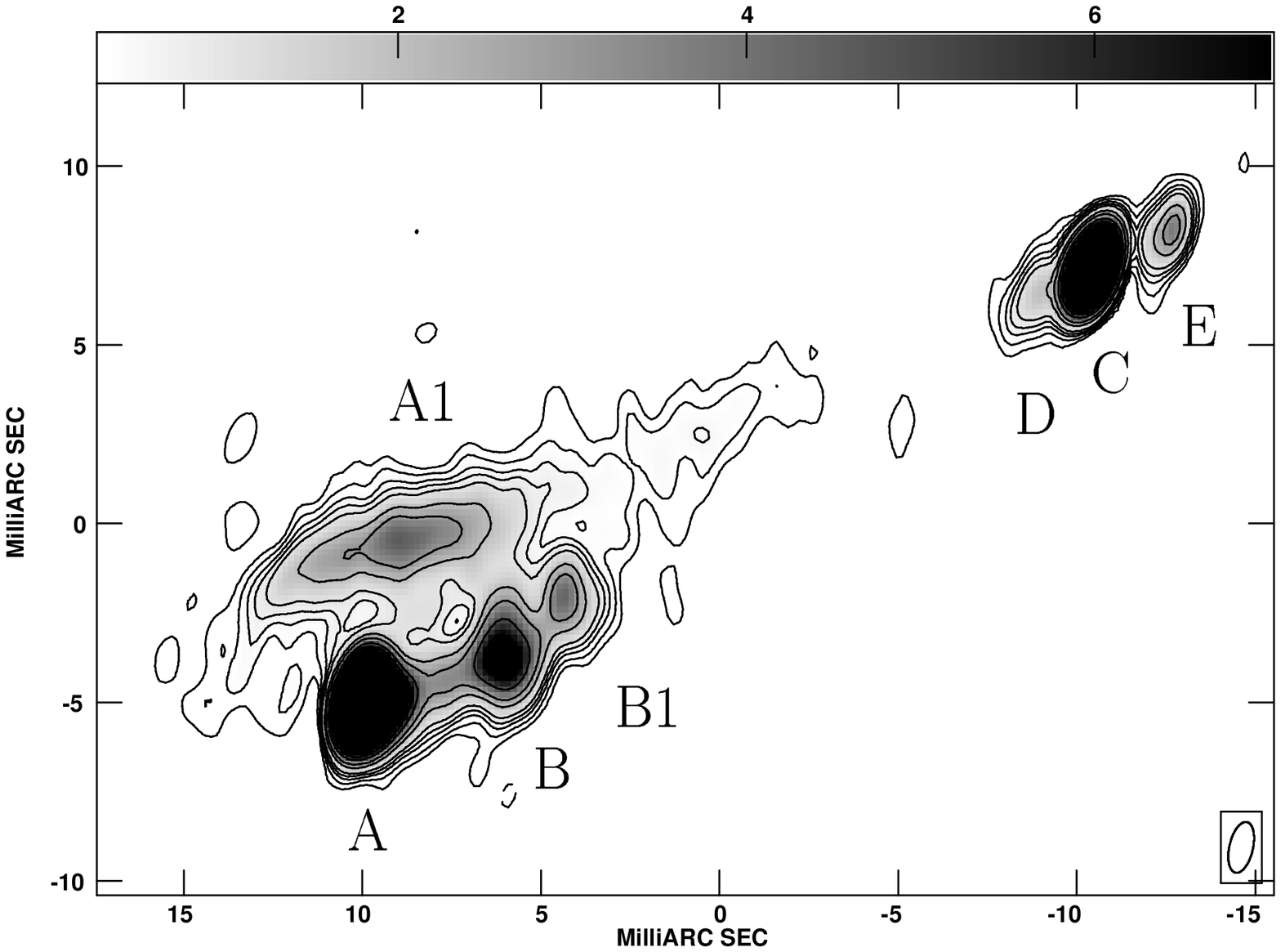} &
\includegraphics[width=3.5in]{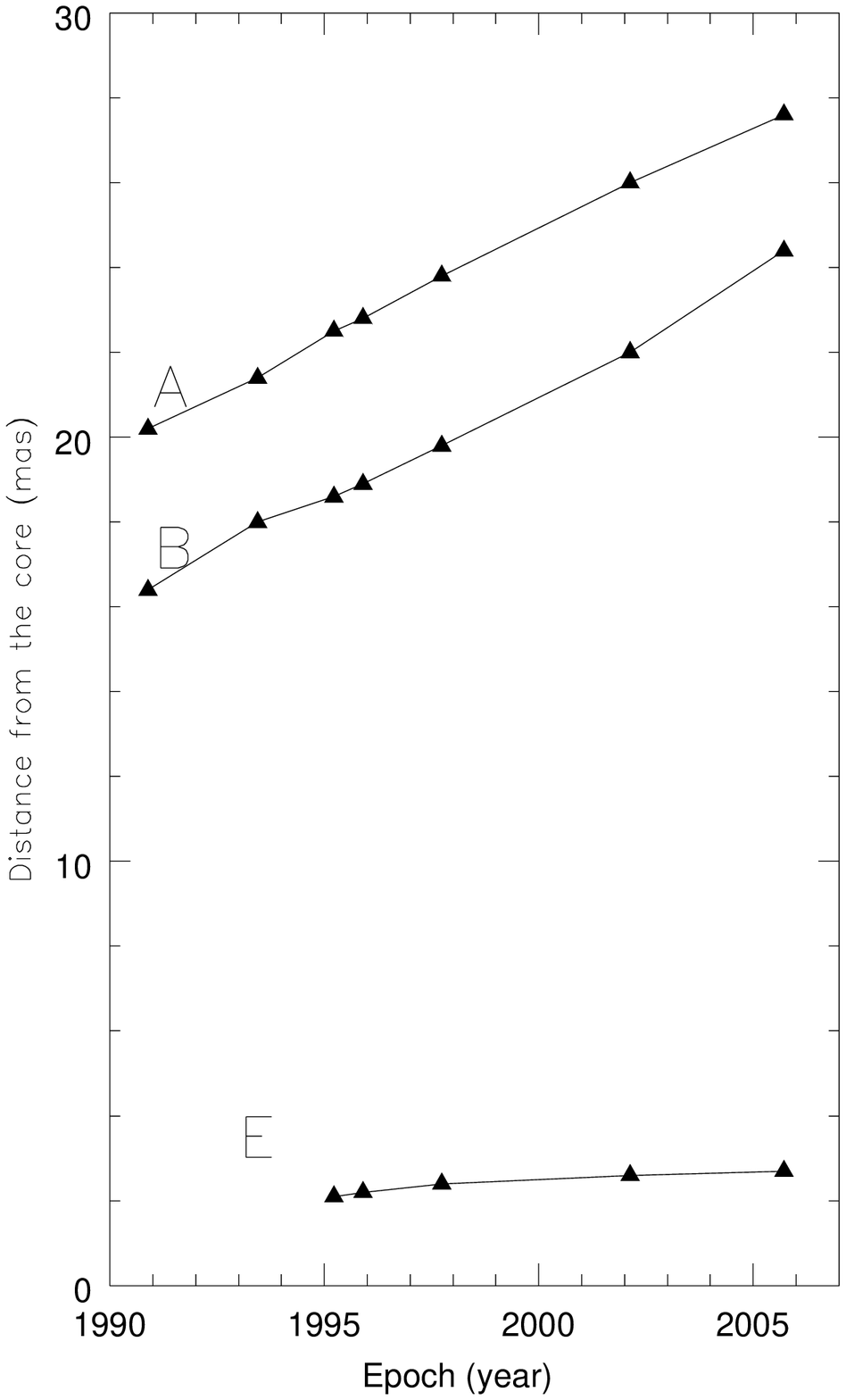} \\
a.~~ VLBI at 8.4 GHz & b.~~  Proper Motion.
\end{tabular}}
\caption{VLBI image of the pc scale structure of 1144+35 (a).
Core distance of components A, B, and E at different epochs (b).}
\label{fig6}
\end{figure}

We used the observations made at different epochs to measure the apparent
proper motion of components A, B, and E with respect to the core
component C. 
In Fig. 6b we show the distance of components A, B, and E from C at
different epochs. 
From these data we measured for the first time a proper motion
of the counter-jet (E) structure (in agreement with the previous upper limit):
$\beta_{a-cj}$ = 0.23 while for components A and B we confirm (note the
different cosmology used here with respect to previous papers):
$\beta_{a-j}$ = 1.92.

Since we know the source Angular Distance (D) in the present cosmology we
can also derive the intrinsic jet orientation \cite{mir99}:

\begin{equation}
\centerline{D = 0.5c tan$\theta$($\mu_a$ - $\mu_r$)($\mu_a\mu_r$)$^{-1}$}
\end{equation}

where $\mu$ is the angular velocity of the approaching (a) and receiding (r) 
jet.

We find $\theta$ = 33$^\circ$ and therefore $\beta$ = 0.94
in agreement wth the measured
jet -- counterjet arm ratio $\sim$ 10 which implies 
$\beta cos\theta$ $\sim$ 0.82
with the possible values: $\beta$ = 0.95, $\theta$ = 30$^\circ$.

The morphology of this source suggests a recurrent radio activity. We have:

\begin{itemize}

\item
The Mpc scale (the oldest structure)
\item
The naked jets and the dominant core on the arcsecond scale
\item
The VLBI structure discussed here.
\end{itemize}

Assuming a constant jet velocity the main VLBI jet structure was emitted from 
the core about on 1950.

One more evidence of a recurring activity is from the arcsecond core
and mas structure flux density variability (Fig. 7).
From the comparison of the light curves of the arcsecond scale core and
the mas scale components it is clear that: 1) the arsecond scale variability 
in not due to the mas core C but to the main jet component (A); 2) in the
last few years (after 2002) the flux density of the mas core C is increasing
and the increase of the flux density is visible also from the arcsecond core 
monitoring but only at high frequency.

\begin{figure}[H]
\centerline{%
\begin{tabular}{c@{\hspace{0.5pc}}c}
\includegraphics[width=3.5in]{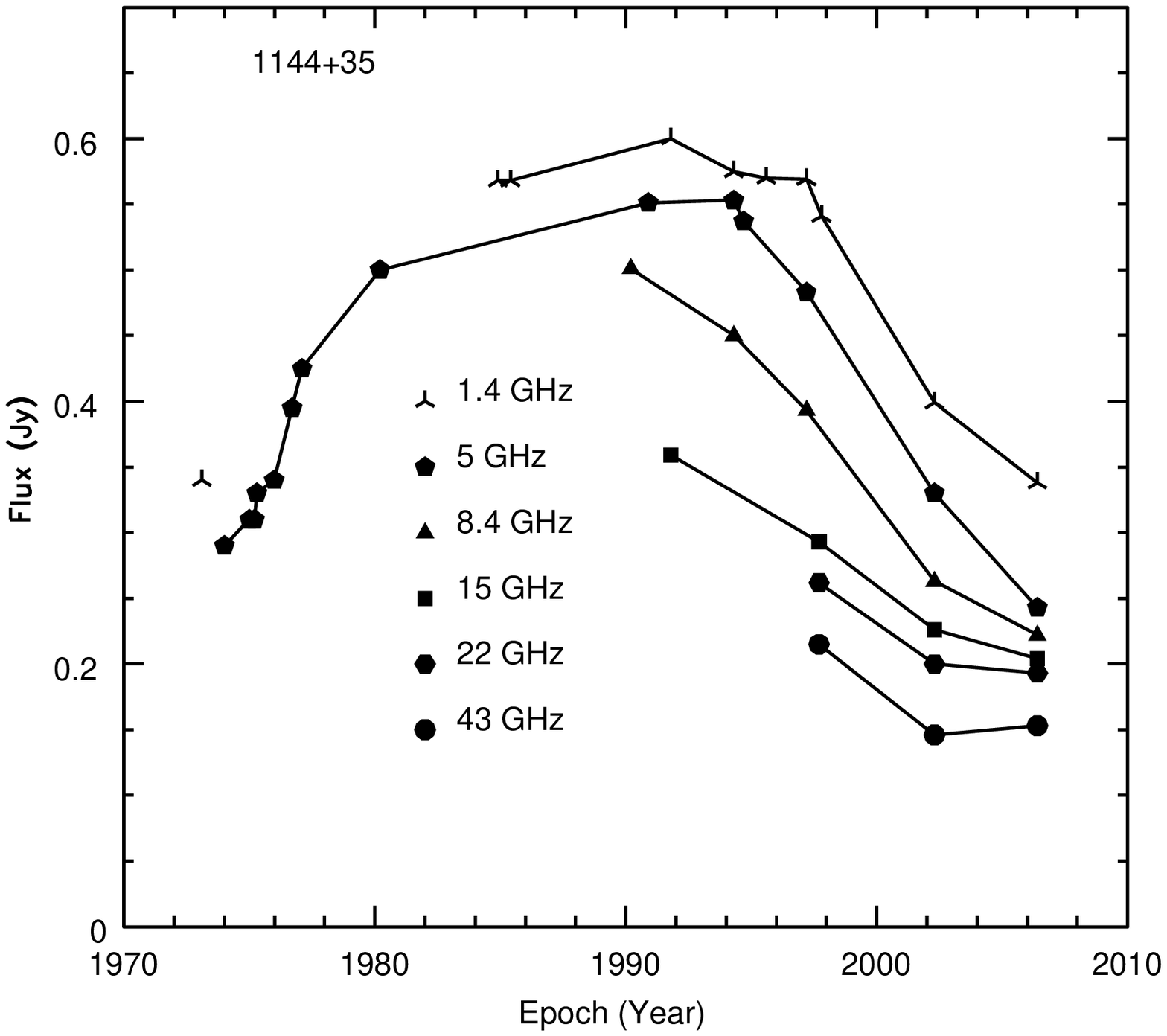} &
\includegraphics[width=3.0in]{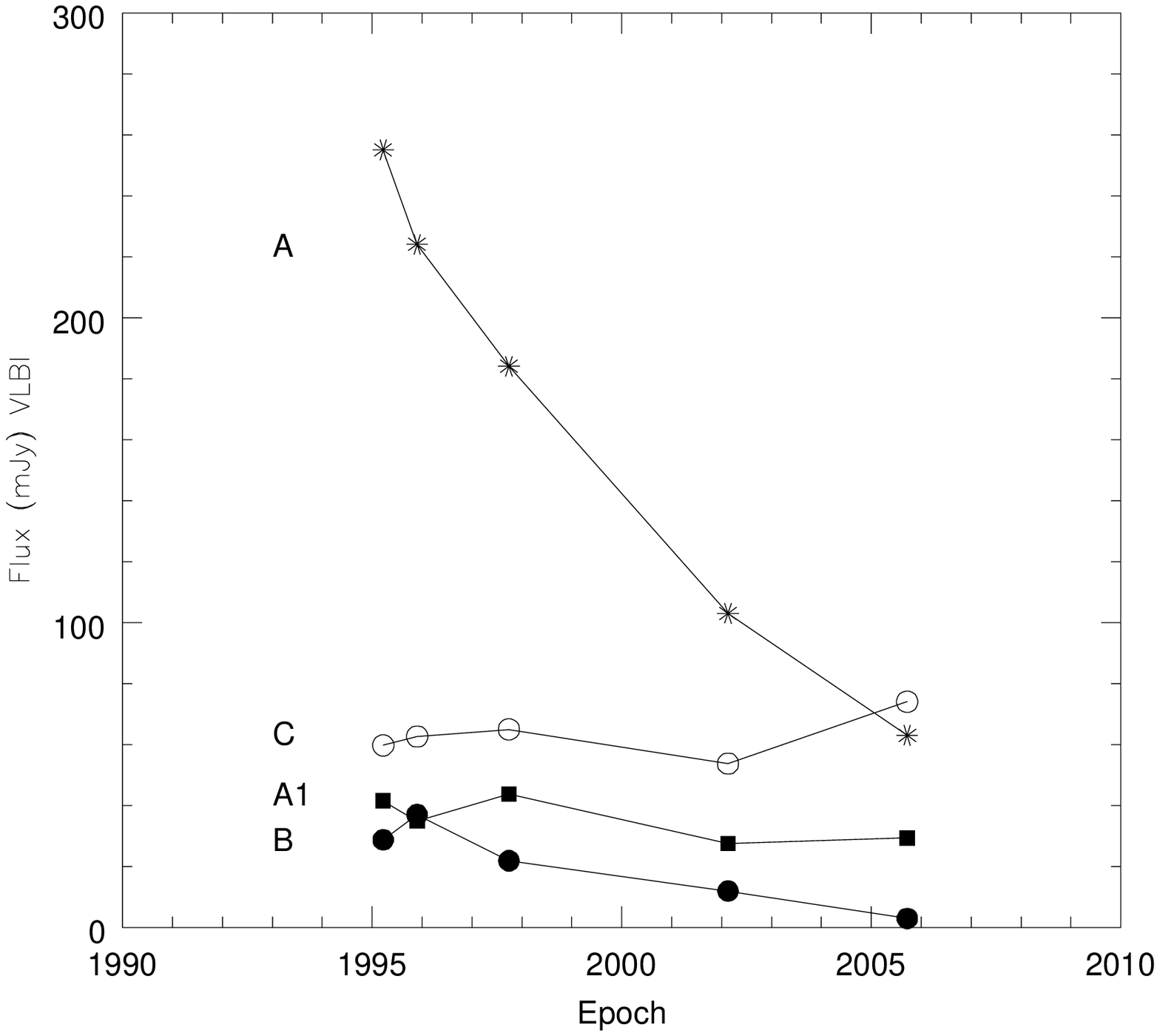} \\
a.~~ Arcsecond core & b.~~  VLBI components.
\end{tabular}}
\caption{Flux density measures of the 1144+35 arcsecond core at different
epochs and frequencies (a).
8.4 GHz flux density measures of components A, A1, B, and C at
different epochs (b).}
\label{fig7}
\end{figure}

A possible explanation is that after 2002 and before of 2006 the core C 
started 
a new active phase and a new component is coming out. This new component
is not yet visible in our images and it is still self-absorbed at frequencies
lower than 8.4 GHz which implies in equipartition conditions a size of this 
new component of about 0.03 mas. If this component is moving at the same 
velocity of the main jet we should start to see it in VLBI images at 8.4
GHz in about 2 years.

\section{Conclusions}

\begin{itemize}

\item
The parsec scale jet morphology is the same in high (FR II) and low 
(FR I) power sources; the pc scale morphology is in agreement with 
expectations from unified models

\item
There is a good agreement between the pc and kpc scale orientation

\item
The pc scale jet velocity is highly relativistic in FR II and FR I sources. 
It is not related to the total or core radio power of the source. 
No correlation was found with the kpc scale structure

\item
In some sources with a low power nuclear source we find a 
 peculiar morphology: restarted activity and a complex mas scale
 structure not yet understood, misaligned with the kpc scale 
 structure.  

\item
The low power source 1144+35 shows a superluminal motion and
 a restarted activity. The core flux density is variable and an
  increase of the core flux density at high frequency suggests
 the presence of a new component which could be visible in 
 VLBI images in 2007-2008

\end{itemize}

\end{document}